\documentclass[
12pt,
prb,
a4paper,
amsmath,
superscriptaddress,
showpacs, 
showkeys,
floatfix,
]{revtex4-1}

\newlength{\figwidth}

\newif\ifOnecolumn
  \ifx\Onecolumn\undefined
  \Onecolumntrue
  
\else
  \Onecolumnfalse
\fi

\newif\ifTwocolumn
  \ifOnecolumn
   \Twocolumnfalse
\else
  \Twocolumntrue
\fi

\usepackage{graphicx,amssymb,amsmath,amsfonts}
\usepackage{epsfig}
\usepackage[mathscr]{eucal}
\usepackage{comment}
\usepackage{bm}

\let \ShowFixme = 1

\begin{document}

\title{Accelerating charging dynamics in sub-nanometer pores}

\author{Svyatoslav Kondrat}
\email{s.kondrat@fz-juelich.de}
\affiliation{Forschungszentrum J\"{u}lich, IBG-1: Biotechnology, 52425 J\"{u}lich, Germany}
\affiliation{Department of Chemistry, Faculty of Natural Sciences, Imperial College London, SW7 2AZ, UK}

\author{Peng Wu}
\author{Rui Qiao}
\email{rqiao@clemson.edu}
\affiliation{College of Engineering and Science, Clemson University, Clemson, South Carolina 29634-0921, United States}

\author{Alexei Kornyshev}
\email{a.kornyshev@imperial.ac.uk}
\affiliation{Department of Chemistry, Faculty of Natural Sciences, Imperial College London, SW7 2AZ, UK}

\begin{abstract}

Having smaller energy density than batteries, supercapacitors have exceptional power density and cyclability. Their energy density can be increased using ionic liquids and electrodes with sub-nanometer pores, but this tends to reduce their power density and compromise the key advantage of supercapacitors. To help address this issue through material optimization, here we unravel the mechanisms of charging sub-nanometer pores with ionic liquids using molecular simulations, navigated by a phenomenological model. We show that charging of ionophilic pores is a diffusive process, often accompanied by overfilling followed by de-filling. In sharp contrast to conventional expectations, charging is fast because ion diffusion during charging can be an order of magnitude faster than in bulk, and charging itself is accelerated by the onset of collective modes. Further acceleration can be achieved using ionophobic pores by eliminating overfilling/de-filling and thus leading to charging behavior qualitatively different from that in conventional, ionophilic pores.

\end{abstract}


\date{\today}

\maketitle

\section{Introduction}

Supercapacitors offer unique advantages of high power density and extraordinary cyclability but provide moderate energy density.\cite{conway:99} Enhancing their energy density without compromising the mentioned advantages would enable their widespread applications.\cite{miller:sci:08} The current surge of interest in supercapacitors is driven by recent breakthroughs in developing novel electrode materials and electrolytes.~\cite{simon:nm:08} In particular, electrodes featuring sub-nanometer pores and room-temperature ionic liquids (RTILs) are among the most promising materials for next-generation supercapacitors: The former affords large specific surface area and may also enhance the specific capacitance\cite{gogotsi:sci:06,gogotsi:08} and energy density;\cite{kondrat:ees:12} the latter allows increasing the operation voltage beyond that of organic electrolytes.\cite{ohno:book,ilSuper2} These materials have enabled impressive improvement of energy density,\cite{ilSuper,ajayan,zhu,subnanoRTIL} and the thermodynamics of charge storage in these materials are now understood reasonably well.~\cite{huang,shim:10, skinner, kondrat:jpcm:11, *kondrat:jpcm:13, wu:ascnano:11,  feng:jpcl:11, jiang:nanolett:11, merlet:natmat:12, xing:jpcl:13} An emerging issue of these materials, however, is that they tend to lower the power density of supercapacitors.\cite{mrsReview} For example, ion transport in RTILs is slow in the bulk and can be even slower in nanoconfinement,\cite{monk:jpcc:11, rajput:jpcc:12, li:langmuir:13, susanReview} leading to sluggish charging dynamics and thus low power density. Resolving these issues, e.g., by judicious selection of pores and RTILs, necessitates a fundamental understanding of the charging dynamics of sub-nanometer pores with RTILs.

The latter is, however, complicated by unique features emerging in sub-nanometer pores. In these pores, all ions of RTIL are in close contact with each other. Consequently, charging dynamics are affected by a multitude of collective effects that cannot be described by existing theories proved valid for mesoporous electrodes, such as the classical transmission line model. Furthermore, conventional ion transport theories, rigorous in the limit of weak ion-ion correlations, cannot be directly used to predict transport of RTILs in nanopores.\cite{bazant:pre:04, biesheuvel:pre:10} 

The comprehensive picture of the charging dynamics of supercapacitors with sub-nanometer porous electrodes and RTILs should answer the following questions:

\begin{itemize}
	\item Are there universal features of charging dynamics in such systems?
	\item Does the slow ion transport in bulk RTILs necessarily imply slow ion transport \emph{during charging} of sub-nanometer pores?
	\item Is it feasible to accelerate charging by tailoring the size, geometry, and surface properties of pores?
\end{itemize}

\begin{figure}[!t]
    \begin{center}
        \ifOnecolumn
    	   	\includegraphics*[width=\textwidth]{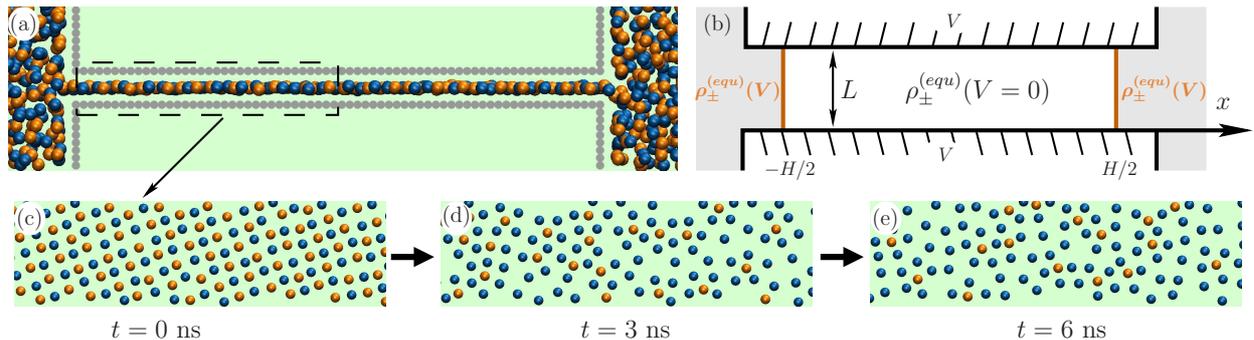}
	\else
		\includegraphics*[width=0.2\textwidth]{figs/model}
	\fi
        \caption{
        \label{fig:model}
	Charging of narrow electrode pores with RTILs. (a) Side-view snapshot of one half of the MD system featuring an electrode pore (width: $0.53$ nm) and part of the RTIL reservoirs connected to it. (b) Schematic of a continuum phenomenological model. The equilibrium ion densities at zero electrode polarization (corresponding to the potential of zero charge) are set inside the pore of width $L$ and length $H$. The ion densities corresponding to a non-zero voltage are set close to the pore entrances, and are kept fixed as the system is let evolve in time. (c-e) Top-view snapshots of the ionic structure inside the negative electrode pore when a voltage of $3$V is imposed impulsively between the positive and the negative electrode pores (for a $5$ns video, see movie M2 in Supplementary Information). The blue and orange spheres represent cations and anions, respectively. Wall atoms are not shown for clarity.}
\end{center}
\end{figure}

Resolving these issues can shape and guide the development of novel materials for supercapacitors. Here we use Molecular Dynamics (MD) simulations and a recently developed~\cite{kondrat:jpcc:13} phenomenological mean-field type (MFT) model to study the dynamics of charging ultrananoporous electrodes with RTILs. We investigate the charging of a pair of slit nanopores in two metallic electrodes, which mimics the nanopores in graphene-based nanoporous electrodes.\cite{ajayan,subnanoRTIL} Figure~\ref{fig:model} shows our nanopore system together with a few snapshots of the time evolution during charging. In MD simulations, we consider ions as charged van der Waals particles of identical size (see Methods). Such approach does not take into account electronic structure of carbon electrodes, neither goes into the details of atomistic structure of real ions. With these deliberate simplifications we aim at revealing the essential physics responsible for generic features of the charging dynamics, unobscured by the chemical complexity of RTILs and real carbon materials. Insights gained from this study will help guide future study of charging dynamics in more complicated situations, e.g., in pores that can accommodate a few layers of ions and in interconnected nanopore networks.  

\section{Charging of nanopores pre-wetted by RTILs}

\subsection{MFT predictions}. 

The flux (along the pore) of monovalent ions confined inside a metallic pore of width comparable to the ion diameter can be written as\cite{kondrat:jpcc:13}
\begin{align}
\label{eq:J}
	J_\pm = - D_\pm \nabla \rho_\pm \mp D_\pm \rho_\pm G \nabla c - \frac{D_\pm \rho_\pm}{\rho_{max} - \rho_\Sigma} \nabla \rho_\Sigma,
\end{align}
where $D_\pm$ is ion's diffusion coefficient (for simplicity we shall use the same $D\equiv D_\pm$), $\rho_\pm$ is the ion density,  $c=\rho_+ - \rho_-$ is the charge (in units of the elementary charge) and $\rho_\Sigma$ the total ion density, and $\rho_{max}$ is the total ion density at close packing; we never reach $\rho_{max}$ in our calculations. $G$ is a parameter characterizing the \emph{screening} of the ion-ion electrostatic interactions due to the electronic polarizability of metallic pore walls; when the pore is made narrower, the screening becomes stronger and $G$ decreases (see Methods). The first term in the ion flux is simply diffusion. The second term comes from the ion `migration'. It is due to the screened electrostatic interactions and is collective in nature. The last term has entropic origin and represents the transport of ions due to the gradient of total ion density along the pore. Equation (\ref{eq:J}) together with the local conservation law define the MFT model for the dynamics of pore charging. 

The RTIL reservoir is not explicitly accounted for in the MFT model. Rather, the ion densities close to the pore entrance are set to the equilibrium densities corresponding to some non-zero voltage (see Methods), and the ion densities inside the pore are let evolve from their equilibrium values at the potential of zero charge (PZC).

\begin{figure}[!t]
    \begin{center}
        \ifOnecolumn
    	   	\includegraphics*[width=\figwidth]{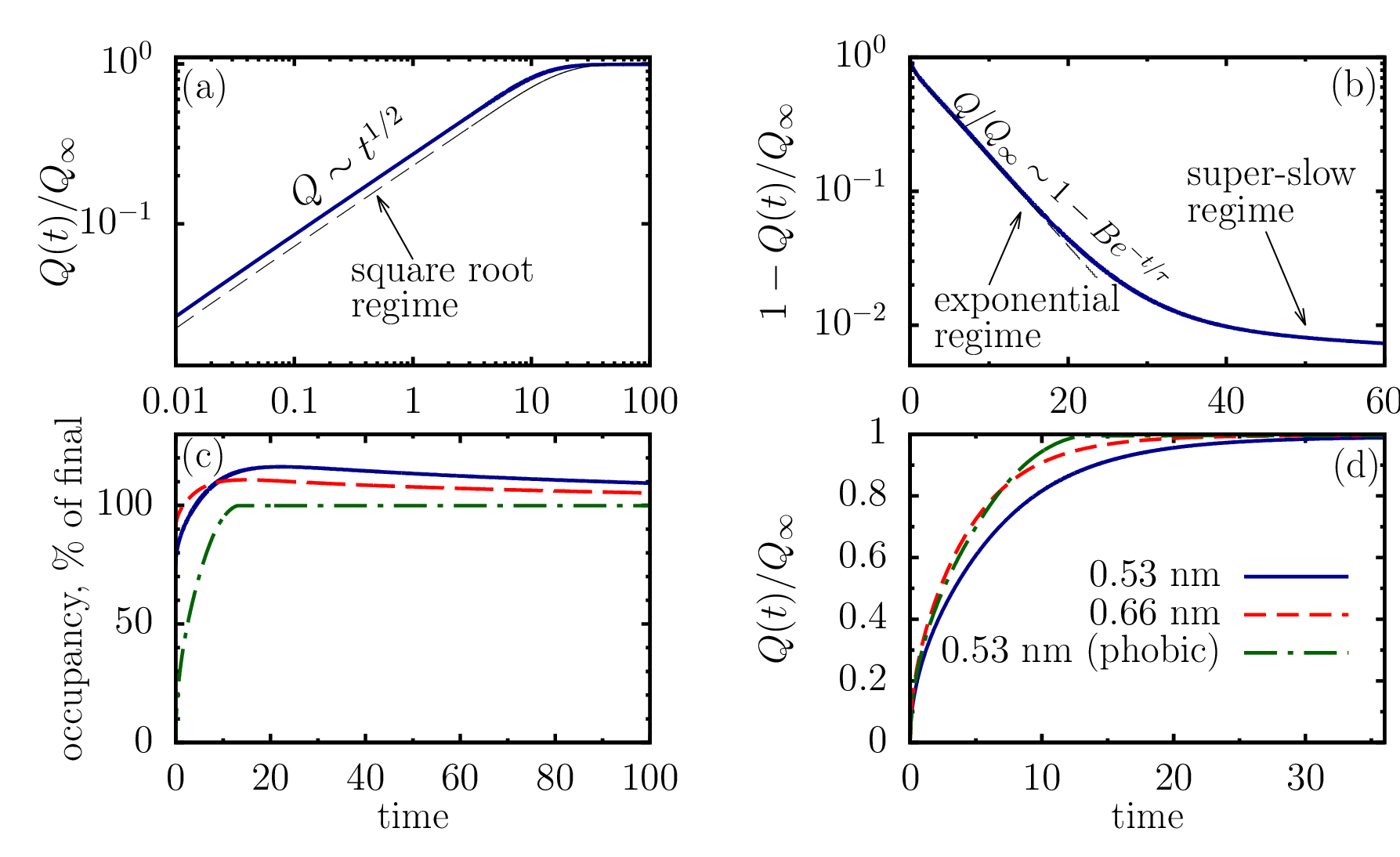}
	\else
		\includegraphics*[width=\figwidth]{mft}
	\fi
        \caption{
        \label{fig:mft}
	Charging of nanopores predicted by the MFT theory. Till a pore is nearly fully charged, $Q(t) = Q_\infty = Q(t=\infty)$, charging is a diffusive process that follows the square root law at short times (panel (a)) and the exponential law at larger times (panel (b)). Charging is accompanied by overfilling (panel c) and is followed by de-filling during which charging is slower. The thin lines in (a) and (b) show the solution of the diffusion equation with $D_{eff}/D \approx 26$, corresponding to zero voltage, in (a), and $D_{eff} \approx 35$ corresponding to the final state ($\approx 0.78$~V) in (b). A long tail in panel (b), deviating from the exponential regime, marks a super-slow charging regime (c.f.~Fig.~\ref{fig:filled:md}(d)). The speed of charging can be improved by making the pore wider or ionophobic (panel (d)). In all plots time is measured in units of $d^2/D$ where $D$ is ion's diffusion coefficient and $d$ its diameter.
	}
\end{center}
\end{figure}

The numerical solution reveals that charging of pores wet by RTILs at PZC is a diffusive process. This can also be seen analytically by noting that the time/space variation of total ion density is small comparing to the variation of charge density (this is true up to times $\approx 15$ in dimensionless units, see movie M1 in Supplementary Information). Then the last term in Eq.~(\ref{eq:J}) can be ignored and one easily arrives at the diffusion equation for the charge density, $\partial_t c = \partial_x D_{eff} \partial_x c (x,t)$, where $D_{eff} (\rho_\Sigma) = D (1 + \rho_\Sigma G)$ is the \emph{effective}  diffusion coefficient. From the analytical solution of this equation\cite{whitaker:book:77} one readily finds the square root behavior at short times (Fig.~\ref{fig:mft}(a))
\begin{subequations}
\label{eq:asymptotics}
\begin{align}
	\label{eq:asymptotics:square}
	Q/Q_\infty \approx 4 \left(D_{eff}/\pi H^2\right)^{1/2} \sqrt{t},
\end{align}
and the exponential saturation at long times (Fig.~\ref{fig:mft}(b))
\begin{align}
	\label{eq:asymptotics:exp}
	Q/Q_\infty \approx 1 - \frac{8}{\pi^2} e^{-t /\tau}
\end{align}
with the relaxation time $\tau = H^2/ \pi^2 D_{eff}$, where $H$ is the pore length.
\end{subequations}

Diffusive nature of charging originates from the fact that the ion migration is proportional to the charge density gradient (see second term in Eq.~(\ref{eq:J}) which follows from the solution of the Poisson equation for the electrostatic potential inside the pore). This contribution enhances the ion transport, as compared to ion's self-diffusion, and leads to $D_{eff}/D \gg 1$. By narrowing the pore, the ion-ion interactions become more screened, thus $G$ and $D_{eff} \sim 1 + \rho_\Sigma G$ decrease; this means that wider pores charge faster (Fig.~\ref{fig:mft}(d)). Interestingly, a similar diffusion slow-down is observed in micellar systems, where the `apparent' diffusion coefficient decreases with adding salt.\cite{galantini:jcp:03} Similarly to our case, where the screening is due to metallic pore walls, the salt screens the electrostatic interactions between the micelles and reduces their collective diffusivity.

The pore occupancy (i.e., the total number of ions inside the pore) increases in the course of charging and reaches values higher than the final, equilibrium occupancy. This \emph{overfilling} is more distinct for narrow pores (Fig.~\ref{fig:mft}(c)) and disappears for sufficiently wide pores (not shown). Interestingly, de-filling extends over time scales much longer than overfilling and is accompanied by a third 'super-slow' regime (c.f.~the long tail in Fig.~\ref{fig:mft}(b)). This super-slow regime, however, seems to be of little practical importance in the present system as the pore is $\approx 99\%$ charged at its onset.

\subsection{Ion diffusion in charged nanopores.} 

Although ions' self-diffusion coefficient is frequently assumed constant,\cite{bazant:pre:04, kilic:pre:07a, *kilic:pre:07b, biesheuvel:pre:10, kondrat:jpcc:13} it depends on ion densities, pore size and other factors. In bulk and in mesopores such dependence is relatively weak or moderate,\cite{rajput:jpcc:12, iacob:sm:12} and can be neglected in many relevant situations. As we shall see, however, this is not the case for sub-nanometer pores, where the ion diffusivity depends dramatically on ion concentrations or degree of pore charging.

\begin{figure}[!t]
    \begin{center}
        \ifOnecolumn
    	   	\includegraphics*[width=\figwidth]{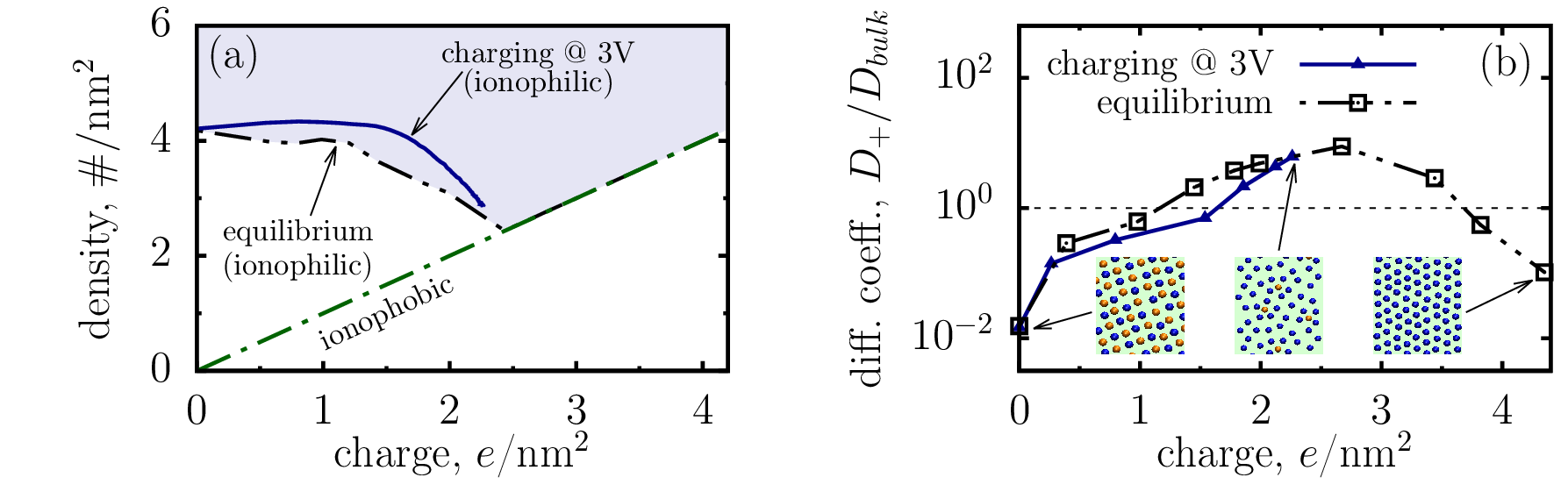}
	\else
		\includegraphics*[width=\figwidth]{md_self}
	\fi
        \caption{
        \label{fig:filled:Dself}
	(a) A map showing the average total ($\rho_\Sigma$) and charge ($c$) densities during `impulsive' charging  (solid line) and in equilibrium (dash double dot line); the dash-dot line corresponds to an iono-phobic pore (c.~f.~Fig.~\ref{fig:empty}). The average total ion density during charging lies within the shaded (blue) area. (b) Cation's self-diffusion coefficient in a $0.53$nm wide pore along the equilibrium path and along $\rho_\Sigma(c)$ corresponding to the impulsive charging at $3$V. The diffusion coefficient is expressed in terms of the diffusion coefficient of a neutral bulk system ($D_{bulk}$). For the $3$V charging, the data only up to $6$ns is shown. Blue and orange spheres in the inset denote the cations and anions, respectively. }
\end{center}
\end{figure}

For other parameters kept fixed, the self-diffusion coefficient ($D_\pm$) turns a complicated function of total ($\rho_\Sigma$) and charge ($c$) densities. For simplicity, therefore, we look at $D_\pm$ along certain `paths' on the $(\rho_\Sigma, c)$ plane, closely related to the actual charging conditions; figure~\ref{fig:filled:Dself}(a) shows such paths. We find that the average total ion density during charging, $\bar \rho_\Sigma (t)$, does not drop below the equilibrium density, $\rho_\Sigma^{(equ)} (c)$, at the same degree of charging, i.e.~for $c = \bar c(t)$. Therefore, we calculate (see Methods) the in-plane self-diffusion coefficient along the equilibrium path, as a limiting case, and compare it with $D_\pm$ along the average $\bar \rho_\Sigma (c)$ at $c = \bar c(t)$ corresponding to the impulsive charging at $3$V. We focus on the diffusion coefficient of cations ($D_+$) and note that $D_-$ shows similar behaviour (see Figs.~S2-S4 in Supplementary Information SI.2). 

The ion diffusion coefficient varies non-monotonically with the charge density inside the pores: When the pore is neutral, ion's self-diffusion is nearly two orders of magnitude slower than in bulk; as the charge inside the pore increases, ion's self-diffusion accelerates and can become $10$ times faster than in bulk; it slows down only when the pore become highly charged ($c \gtrapprox 2.6 e$/nm$^2$). These phenomena seem general and are observed in equilibrium and during `impulsive' charging, for wider pores and for more realistic RTILs (Figs.~S2-S4 in SI.2).

The non-monotonic variation of the diffusion coefficient originates from the different structure of an ionic liquid inside the pore at different states of charging (see insets in Fig. \ref{fig:filled:Dself}(b)). At PZC, ions form a two-dimensional lattice with counter- and co-ions interlocked with each other like in an ionic crystal. Diffusion of ions in such an environment requires large activation energy to unbind counter/co-ion pairs\cite{klahn:jpcb:08} or to cleave their `bonds', and thus the ion diffusion is slow. As more counter-ions are introduced, the perfect inter-locked counter/co-ion lattice gradually disappears and ions diffuse more freely. Such accelerated self-diffusion has also been observed near charged planar surfaces\cite{rajput:jpcc:12}, but the effect is moderate. This is because counter-ions near charged surfaces are still bounded to many co-ions in adjacent ionic layers. When the ions form a monolayer inside a narrow pore, such binding disappears and the acceleration of ion diffusion is much more dramatic. At large counter-ion density, ions form a quasi-Wigner crystal with a small number of co-ions as impurities, and the diffusion coefficient decreases. In this case, however, there is mostly steric contribution to the activation energy, which is much lower than at PZC, and hence the diffusion in highly charged pores is much faster than at PZC.

We thus conclude that a careful examination of RTILs inside nanopores precisely under charging conditions is necessary for selecting an optimal electrode/RTIL pair, rather than a simple `extrapolation' of RTIL's bulk properties. While this renders the design of RTILs more complex, it also opens up exciting opportunities for tailoring RTILs for specific pores and degrees of charging.

\subsection{Charging Dynamics from MD simulations.} 

Let us now return to the dynamics of charging. We impose `impulsively' a potential difference of $3$V between the negative and positive electrodes, analyze however the charging of only one electrode pore (negative, to be specific, see Fig.~\ref{fig:model}(a)), as our system is fully symmetric; we shall also restrict our considerations to pores of two different widths.

\begin{figure}[!t]
    \begin{center}
        \ifOnecolumn
    	   	\includegraphics*[width=\figwidth]{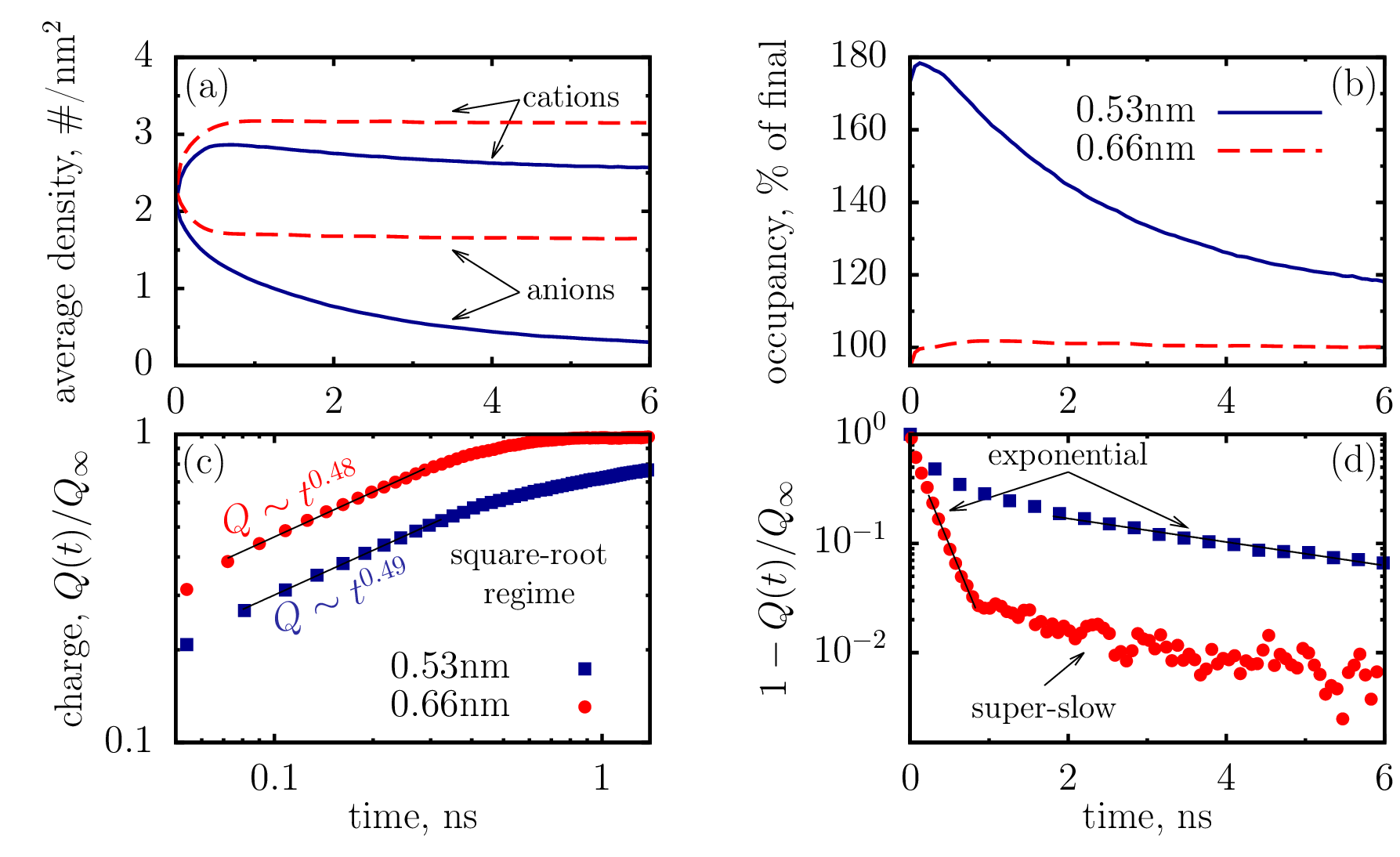}
	\else
    	   	\includegraphics*[width=\figwidth]{md_philic}
	\fi
        \caption{
        \label{fig:filled:md}
	Charging of ionophilic pores of length $12.09$ nm obtained from MD simulations. A voltage of 3 V is imposed impulsively between the negative and positive electrodes at $t=0$. The average cation and anion densities are shown in panel (a) and the total pore occupancy in (b). The evolution of the net charge inside the pores exhibits a diffusive behavior: the initial stage of charging follows a square-root law (c) and the late stage of charging follows an exponential saturation law (d).}
\end{center}
\end{figure}

In line with the MFT predictions, the pore occupancy behaves non-monotonically with time. Initially, the incoming flux of counter-ions overweights the outgoing flux of co-ions, leading to a slight overfilling (Fig.~\ref{fig:filled:md}(a) and (b)). Although overfilling is similar for both pores, the subsequent de-filling differs significantly. In case of a wider pore ($0.66$nm), the pore occupancies at PZC and in the final state (corresponding to $3$V) are comparable, and de-filling has little effect on charging. Indeed, we find that the accumulated charge reaches nearly $98\%$ of the final charge at the onset of de-filling, which is thus mainly characterized by `removal' of both co- and counter-ions from the pore. This is followed by a 'super slow' regime similar to the one predicted by the MFT (see Fig.~\ref{fig:mft}(b)). In the narrower pore ($0.53$nm), the difference between the initial (at PZC) and final occupancies is considerable, and the charging in later times is essentially due to de-filling (see video M2 in Supplementary Information), which leads to a significant slow-down of the charging process.

The evolution of net charge inside the pore, $Q(t)$, exhibits the square-root and exponential saturation regimes revealed by the MFT model (compare Figs.~\ref{fig:mft}(a-b) and Figs.~\ref{fig:filled:md}(c-d)). Motivated by this, we use Eqs.~(\ref{eq:asymptotics}) to fit $Q(t)$ and extract the effective diffusion coefficients, $D_{eff}$; note that $D_{eff}$ characterizes the whole system in a given time frame. In the square-root regime, Eq.~(\ref{eq:asymptotics:square}), we get $D_{eff} = 3.09 \pm 0.39 \times 10^{-9}$~m$^2/$s for $L=0.53$~nm and $D_{eff} = 7.17 \pm 0.88 \times 10^{-8}$~m$^2/$s for $L=0.66$~nm wide pore; in the exponential regime, Eq.~(\ref{eq:asymptotics:exp}), we obtain $D_{eff} = 0.4 \pm 0.08 \times 10^{-8}$~m$^2/$s and $D_{eff} = 5.91\pm 0.71 \times 10^{-8}$~m$^2$/s, respectively. The extracted values of $D_{eff}$ show a decrease with reducing pore width, manifesting slower charging in narrower pores. 

It is instructive to compare $D_{eff}$ with the self-diffusion coefficient ($D_\pm$). This is impeded however by the fact that $D_\pm$ varies with RTIL density and composition (recall however that $D_+ \approx D_- \approx D$, see Figs.~S2-S4 in SI.2). To be on a safe side, in most cases we take the highest value of $D$ at relevant conditions (see section SIII.B in SI.2). For the $0.66$nm pore we get $D_{eff}/D \approx 30$ in both square-root and exponential regimes. This is in qualitative agreement with the MFT, which predicts a considerable enhancement of ion transport due to collective effects (the second term in Eq.~(\ref{eq:J})) 

Similar enhancement is obtained in the square-root regime for the narrower pore ($0.53$~nm), $D_{eff} / D \approx 10$. At later times, however, $D_{eff}$ becomes comparable to the self-diffusion coefficients, with $D_{eff}/D \approx 0.5 - 1.0$. This is closely related to the de-filling character of charging discussed above. In this case, the first and third terms in Eq.~(\ref{eq:J}) dominate, and charging becomes subdominant to de-filling. Physically, such a slow-down can be understood by noting that low co-ion concentrations and strong screening of ion-ion interactions in nanopores reduce collective effects. In other words, the co-ions have to diffuse on their own in the sea of counter-ions, to find a way out of the pore, and hence $D_{eff}$ becomes comparable to $D$. 

\section{Accelerating charging by engineering nanopore surface properties}

Our results suggest that charging of narrow pores is nearly always accompanied by overfilling, which itself is a fast process. The price one has to pay, however, is de-filling, which turns out to slow down charging significantly. It seems thus beneficial from practical point of view to use electrodes with wide pores, where overfilling and hence de-filling are reduced or vanish. Unfortunately, however, in most cases increasing pore size deteriorates capacitance and stored energy density.\cite{gogotsi:08, kondrat:ees:12}

Motivated by the MFT results (Figs.~\ref{fig:mft}(c-d) and Ref.~\onlinecite{kondrat:jpcc:13}), we explore here a different possibility of accelerating charging, by making the surface of nanopores \emph{ionophobic}. Pore ionophobicity can be achieved, for instance, by using mixtures\cite{lin:jpcl:11} of different RTILs or by adding surfactants.\cite{fic:ea:10, *fic:ea:11} In this work we mimic it by tuning the ion-wall van der Waals interactions, so that the pores are free of RTILs at PZC (see Methods).

\begin{figure}[!t]
    \begin{center}
        \ifOnecolumn
    	   	\includegraphics*[width=\figwidth]{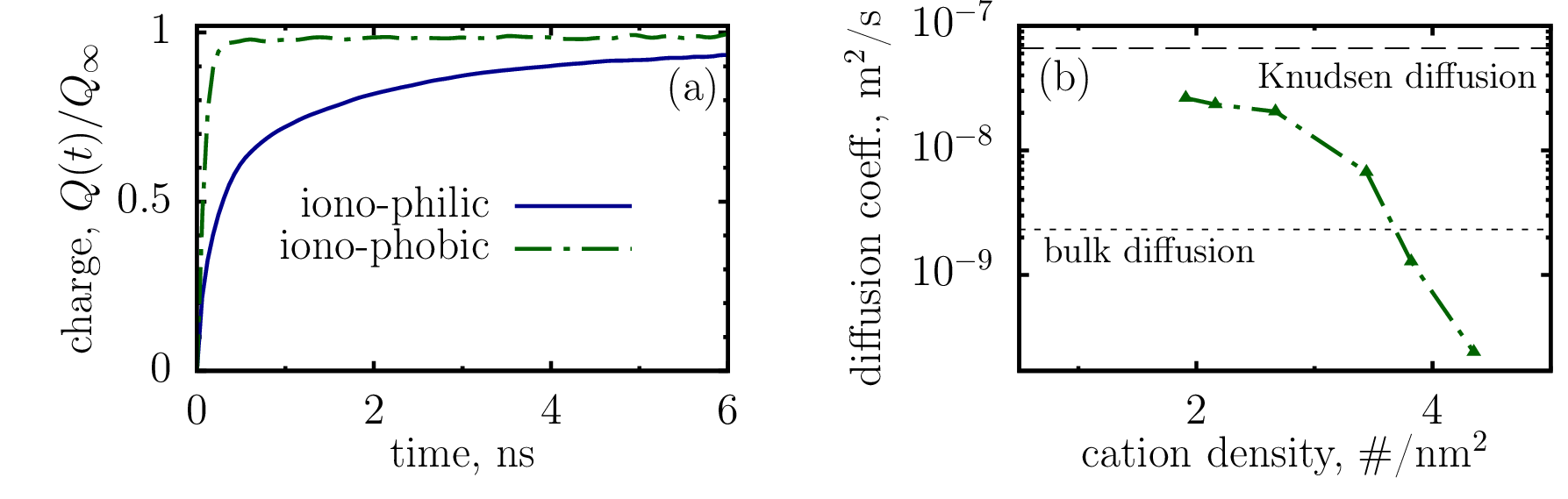}
	\else
		\includegraphics*[width=\figwidth]{md_phobic}
	\fi
        \caption{
        \label{fig:empty}
	 (a) Comparison of charging ionophobic and ionophilic pores of the same width ($0.53$ nm). The ionophobic pore is free of ions at potential of zero charge. (b) Cation's self-diffusion coefficient inside a negatively charged ionophobic pore at different cation densities (no anions present in the pore). The dashed line denotes the Knudsen diffusion coefficient. For comparison, short dash line shows the bulk diffusion coefficient (see Fig.~\ref{fig:filled:Dself}).}
\end{center}
\end{figure}

Ionophobic pores charge initially in a front like fashion, with counter-ions spreading quickly throughout the pore (see video M3 in Supplementary Information); this is followed by a slower `diffusive' like charging, much alike wide ionophilic pores. Importantly, however, we find that  ionophobic pores charge \emph{order of magnitude} faster than ionophilic pores at the same conditions. For instance, in the ionophilic pore $90\%$ of charging is achieved in $4$ns, while only $\approx 0.2$ns is needed in case of an ionophobic pore.

A distinct feature of ionophobic pores is the behavior of self diffusion coefficient ($D_+$ in our case). At early stage of charging, the ion density inside the pore is low and the ion-ion separation is much larger than the ion-wall separation, hence the ion diffusion is limited by collisions with the pore walls. In this case, the self-diffusion coefficient is very large and approaches the Kundsen limit (Fig.~\ref{fig:empty}(b)). As more counter-ions enter the pore, the diffusion coefficient gradually reduces. Importantly, the pore becomes highly charged before the diffusion coefficients decreases significantly. For instance, when charging reaches $90\%$, the self-diffusion coefficient, $D_+ \approx 2.57 \times 10^{-8}~$~m$^2/$s, is higher than in an ionophilic pore and in the bulk at comparable conditions. Incidentally, the strong  variation of $D_+$ explains why the MFT, where we assumed a constant diffusion coefficient, underestimates the acceleration of charging due to ionophobicity of pore walls.

Finally, it is interesting to note that we observe a transition between collective Fickian diffusion and (nearly) self-diffusion in both iono-phobic and -philic pores. Its effect on charging is different, however. For ionophilic pores the charging undergoes a transition from collective to self-diffusion, and this slows down charging. On the contrary, for ionophobic pores a transition from Knudsen type self diffusion to collective diffusion is observed, and the onset of collective modes slows down the dynamics.

\section{Summary}

In summary, a phenomenological model and molecular dynamics simulations show that charging of ionophilic pores, of width comparable to the ion diameter, follows an effective diffusion law. Such charging is a complex process, complicated by a myriad of factors, as extreme confinement and ion crowding, image forces and screened interactions, etc. Thus, the `law of effective diffusion' is not only remarkable but also of practical importance. Indeed, it can for instance help simplify the development of `whole porous-electrode' models, and thus open doors for optimizing electrode materials beyond single-pore level.

Ion's self-diffusion in sub-nanometer pores shows an interesting dependence on ion densities and composition. The self-diffusion coefficient varies during charging over a few orders of magnitude, and can exceed a few times the ion diffusion in the bulk (under similar conditions). This suggests that fast charging can in principle be achieved if an ionic liquid is  optimized specifically for selected porous materials and the required degree of charging.

We have found that charging is often accompanied by overfilling. Although overfilling can in fact accelerate charging, as demonstrated by high effective collective diffusivity, the subsequent de-filling slows down charging significantly, and shall be avoided in practical applications. One way to achieve this is to make pores ionophobic. We have shown that ionophobic pores can accelerate charging by an order of magnitude. Our preliminary calculations and the recent experience with cylindrical pores~\cite{lee:13} show that pore ionophobicity leads to comparable values of capacitance and enhanced energy density at moderately high voltages. We therefore believe that  ionophobic pores present an exciting opportunity for increasing both power and energy density of nanoporous supercapacitors.

\section{Methods}


\subsection{Mean-field model.} 

We consider a single layer of ionic liquid confined in a slit nanopore formed by two parallel metal walls. The free energy of the system can be written as\cite{kondrat:jpcc:13} $F[\rho_\pm] = E_{el}[\rho_\pm] -T S[\rho_\pm] + \int \boldsymbol{(} h_+ \rho_+(x) + h_- \rho_-(x)\boldsymbol{)} dx$, where $T$ is temperature and $\rho_\pm$ ion densities. In the first term we take into account explicitly the pore-induced exponential screening of the ion-ion electrostatic interactions\cite{kondrat:jpcm:11}. To account for excluded volumes, we adopt the Borukhov-Andelman-Orlando expression\cite{borukhov:97} for the entropy, $S[\rho_\pm]$. The voltage-dependent `external fields,' $h_\pm$, consist of ion's electro-chemical potentials, resolvation energy, and the van der Waals and image-force\cite{kondrat:jpcm:11} interactions of ions with the pore walls. The $h_\pm$ control the equilibrium ion densities inside the pores and do not participate in the dynamics other than via initial and boundary conditions.

The dynamics is defined by the continuity equation $\partial_t \rho_\pm = - \partial_x J_\pm$. For the current we postulate $J_\pm = - \Gamma \; \partial_x  (\delta F / \delta \rho_\pm)$, where $\Gamma \equiv \Gamma_\pm = D_\pm / k_BT$ is a phenomenological mobility parameter and $D_\pm$ the diffusion constant, which we assumed pore-width, voltage and density independent; $k_B$ is the Boltzmann constant, as usual. Plugging the free energy $F[\rho_\pm]$ in the continuity equation results in Eq.~(\ref{eq:J}): The first and third terms follow from the entropy, and the second term is due to $E_{el}$, where\cite{kondrat:jpcc:13}
\begin{align}
	G = 4 L_B R_c \; \sum_{n=1}^\infty \frac{\sin^2(\pi n /2)}{n} K_1\left( \pi n R_c/L\right)
\end{align}
is a parameter characterizing screening of the electrostatic interactions by the metal pore walls. Here $K_1$ is the modified Bessel function of the second kind of first order, $R_c$ is the cut-off radius\cite{kondrat:jpcc:13} and $L_B = e^2/\varepsilon_p k_B T$ (in Gaussian units) is the Bjerrum length, where $e$ is the elementary charge and $\varepsilon_p$ the dielectric permittivity inside the pore (we assumed $\varepsilon_p$ pore-width independent; for the effect of pore-width varying dielectric permittivity see Ref.~\onlinecite{kondrat:ec:13}). The $G$ depends on the pore width, $L$, and decreases with narrowing the pore.

Equations~(\ref{eq:J}) were solved numerically using the GSL library\cite{gsl}. The solution provides the ion densities, and thus the accumulated charge and pore occupancy at any given time. 

\subsection{Molecular Dynamics (MD) simulations.} 

The MD system consists of a pair of identical slit pores and two reservoirs separating the pores. The access width of pores was $0.53$nm and $0.66$nm, the pore length was $12.09$nm, and periodic boundary conditions were applied in all directions. Each pore wall was made of a square lattice of Lennard-Jones (LJ) particles, and cations and anions were modeled as charged LJ particles. The ionophobicity of the pore wall was varied by tuning the LJ parameters of the ion-wall interactions. A schematic picture of the MD system and the force field parameters are provided in the Supplementary Information (section SI in SI.2).

MD simulations were performed using a customized Gromacs code.\cite{gromacs} Pore walls were maintained as equi-potential surfaces with their image planes coinciding with the geometric plane of wall atoms. In the method\cite{anjan} we used, the electronic polarizability of pore walls is taken into account on the continuum electrostatics level. This method is in good agreement\cite{pengJPCL,wu:ascnano:11}  with other models of polarizable electrodes.\cite{merlet:natmat:12,xing:jpcl:13} 

The system was first equilibrated for $2$ns at PZC. The number of ions inside the entire system was tuned so that the ion density in the RTIL reservoirs matched that of a bulk system at $400$K and $1$atm (such an elevated temperature was chosen to ensure that the model RTIL remains in the liquid phase). After the system reached the equilibrium, a voltage difference was impulsively imposed between the negative and positive electrodes, and the system was let evolve in the NVT ensemble for $6$ns. Each charging case was repeated $50$ times, with independent initial configurations, to obtain reliable statistics. To compute the equilibrium charge at a given applied voltage, a separate system with $50$\% shorter pores were setup and run for $20$ns. 

To study self-diffusion of ions inside nanopores, we setup MD systems which consist of a single pore only (and the ions in it), with periodic boundary conditions in all directions. We tuned the number of cations and anions to match the desired total and charge densities inside the pore. Since pore walls are modeled as equi-potential surface, they form a Faraday cage around ions, and electroneutrality is automatically satisfied. The diffusion coefficient of ions was computed using the Einstein-Helfand relation\cite{frenkel:book}; the ion trajectories were obtained from at least $5$ns equilibrium runs. 

\begin{acknowledgments}
We thank the Clemson-CCIT office for providing computer facilities. R.Q.~acknowledges the support of the NSF (CBET-1246578). S.K.~and A.K.~were supported by the Engineering and Physical Science Research Council via Grant EP/H004319/1. We are thankful to Yury Gogotsi, Patric Simon, Carlos P\'erez, John Griffin, Gleb Oshanin and Fritz Stoeckli for fruitful discussions, and Xikai Jiang for technical assistance.
\end{acknowledgments}



%

\end{document}